\documentclass[a4paper]{article}

\usepackage{itgspeech2025}    
\usepackage{times}            
\usepackage[english]{babel}   
\usepackage[ansinew]{inputenc}
\usepackage[T1]{fontenc}      
\usepackage[sort&compress,numbers]{natbib}	
\usepackage{amsmath,amssymb}
\usepackage{graphicx}
\usepackage[colorlinks=false,pdfborder={0 0 0}]{hyperref}
\usepackage{units}

\usepackage{amsmath,amssymb,amsfonts}
\usepackage{algorithmic}
\usepackage{graphicx}
\usepackage{textcomp}
\usepackage{xcolor}
\usepackage{orcidlink}
\usepackage{caption}
\usepackage{acronym}
\usepackage{pifont}
\usepackage{booktabs}
\usepackage{balance}

\usepackage{tikz}
\usetikzlibrary{matrix}
\usepackage{pifont}
\usepackage{diagbox}

\newcommand{\yes}{|[fill=green!40]|\ding{52}}
\newcommand{\yesRMSE}[1]{|[fill=green!40]| {\ding{52} \\ (RMSE = #1)}}
\newcommand{\yesQM}{|[fill=green!40]| {\ding{52} \\ (Intended?)}}
\newcommand{\no}{|[fill=red!40]|\ding{56}}

\usepackage{tikz}

\title{Navigating PESQ: Up-to-Date Versions and Open Implementations}

\author{Matteo Torcoli\textsuperscript{\dag}\thanks{M.\,Torcoli and M.M. Halimeh were affiliated with Fraunhofer IIS at the time the work was conducted. M. Torcoli is now with Amplifon and M.M. Halimeh is now with Starkey Hearing Technologies.
\\
\hspace*{1,5em}\textsuperscript{2}A joint institution of Fraunhofer IIS and Friedrich-Alexander-Universit\"at Erlangen-N\"urnberg (FAU), Germany.
}\textsuperscript{\orcidlink{0000-0003-2834-9194}}, Mhd Modar Halimeh\textsuperscript{\dag$1$\orcidlink{0000-0001-6113-3067}}, and Emanu\"el A. P. Habets\textsuperscript{\dag\ddag\orcidlink{0000-0002-2613-8046}}}

\address{\textsuperscript{\dag}Fraunhofer Institute for Integrated Circuits IIS, Erlangen, Germany\\
\textsuperscript{\ddag}International Audio Laboratories Erlangen\textsuperscript{2}, Erlangen, Germany\\
  Email: \texttt{matteo.torcoli@amplifon.com, \\ \{mhd.modar.halimeh,emanuel.habets\}@iis.fraunhofer.de}}

\acrodef{ITU-T}[ITU-T]{International Telecommunications Union -- Telecommunications Sector}
\acrodef{PESQ}[PESQ]{Perceptual Evaluation of Speech Quality}
\acrodef{POLQA}[POLQA]{Perceptual Objective Listening Quality Analysis}
\acrodef{ODAQ}[ODAQ]{Open Dataset of Audio Quality}
\acrodef{RMSE}[RMSE]{Root Mean Squared Error}
\acrodef{DE}[DE]{Dialogue Enhancement}
\acrodef{DNN}[DNN]{Deep Neural Network}
\acrodef{DS}[DS]{Dialogue Separation}
\acrodef{SE}[SE]{Speech Enhancement}

\begin{document}

\maketitle    

\sloppy

\begin{abstract}
\ac{PESQ} is an objective quality measure that remains widely used despite its withdrawal by the International Telecommunication Union (ITU).
PESQ has evolved over two decades, with multiple versions and publicly available implementations emerging during this time.
Different versions and their updates can be overwhelming, especially for new PESQ users.
This work provides practical guidance on the different versions and implementations of PESQ.
We show that differences can be significant, especially between PESQ versions. 
We stress the importance of specifying the exact version and implementation that is used to compute PESQ, and possibly to detail how multi-channel signals are handled. 
These practices would facilitate the interpretation of results and allow comparisons of PESQ scores between different studies.
We also provide a repository that implements the latest corrections to PESQ, i.e., \mbox{Corrigendum~2}, which is not implemented by any other openly available distribution: \url{https://github.com/audiolabs/PESQ}.

\end{abstract}

\section{Introduction}
\subsection{PESQ history, importance, and caveats}
\ac{PESQ} \cite{rix2001perceptual, ITUT:2001, PESQ:682.1, PESQ:682.2, PESQ:c2} was developed to predict the mean opinion score (MOS) of a transmitted speech signal by comparing it with a clean reference. \ac{PESQ} has a long history as an \ac{ITU-T} recommendation P.862, with its first version published in 2001 and the last document dated 2018, as summarized in Table~\ref{tab:docs}.
In 2018, \ac{PESQ} was withdrawn by the ITU to be replaced by \ac{POLQA} \cite{beerends2013perceptual, POLQA:2018}.
Despite its withdrawal, \ac{PESQ} remains a widely used metric for researchers and practitioners. This is evident by conducting a Google Scholar search for the term \textit{PESQ}, which returns over 4,600 search results for the year 2024, while only a little more than 100 results are found for the term \textit{POLQA}.
PESQ has been used as part of the training strategy or validation metrics for \ac{SE} in various works, e.g., \cite{fu2021metricgan,braun2021towards, Xu2022, de2024pesqetarian}, and methods have been proposed to predict PESQ scores non-intrusively (i.e., without a clean reference) using a \ac{DNN} \cite{Catellier2020, Kumar2023}.
The popularity of PESQ is due to several reasons; chief among them is the abundance of publicly available implementations and the robustness that PESQ shows across different application domains, although its correlation with perceived quality is outperformed by other measures \cite{torcoli:2021}.

Given the evolution of PESQ and its versions, the availability of several implementations can lead to confusion and may render the PESQ scores reported in different works incomparable.
This paper acknowledges the importance of PESQ in the \ac{SE} community and aims to provide guidance in navigating its numerous versions and implementations.
This paper neither endorses the use of PESQ nor aims to improve it.
The validity and accuracy of PESQ as a quality metric are beyond the scope of this work.

\subsection{Contributions and structure of this work}
This work examines PESQ implementations and their differences within the larger context of ITU recommendations.
The PESQ versions and implementations considered in this work are summarized in Table~\ref{tab1}.
The remainder of this work is structured as follows. 
Sec.~\ref{sec:ver} describes the different versions of PESQ. Sec.~\ref{sec:imp} discusses the different publicly available implementations. Sec.~\ref{sec:differences} examines the differences between the versions and the implementations, including a discussion of the used test audio signals.
The application to multi-channel signals is discussed in Sec.~\ref{sec:multi}. Section~\ref{sec:conc} concludes this work.

\section{PESQ Versions}
\label{sec:ver}
\subsection{Narrowband raw PESQ (P.862)}
PESQ was designed to evaluate the quality of speech signals transmitted over telecommunication networks.
The method includes a pre-processing step that mimics a telephone handset.
Measures for audible disturbances are computed depending on the loudness of the signals and combined into raw PESQ scores according to P.862 \cite{ITUT:2001}, i.e., the first version of PESQ.

\begin{table}[t]
\small
\centering
    \begin{tabular}{l l}
    \toprule
                            & P.862 (02/2001) \\
                            & Amendment 1 (03/2003) \\
    P.862 \cite{ITUT:2001}  & Amendment 2 (11/2005) \\
                            & Corrigendum 1 (10/2007) \\
                            & Corrigendum~2 (03/2018) \cite{PESQ:c2} \\
    \midrule
    P.862.1 \cite{PESQ:682.1} & P.862.1 (11/2003) \\
    \midrule
                              & P.862.2 (11/2005) \\
    P.862.2 \cite{PESQ:682.2} & P.862.2 (11/2007) \\
                              & Corrigendum 1 (10/2017)\\
    \bottomrule
    \end{tabular}
\caption{List of ITU publications defining the different versions of PESQ. In addition to the listed publications, P.862.3 (2005, 2007, and 2011) provides remarks on how to apply P.862, P.862.1, and P.862.2. The documents can be found at {\small\url{https://www.itu.int/rec/T-REC-P/en}}.}
\vspace{-2em}
\label{tab:docs}
\end{table}

\begin{table*}[hbt!]
\footnotesize
\begin{tikzpicture}
    \matrix (magic) [matrix of nodes,
    nodes={minimum width=2.82cm,text width=2.8cm,minimum height=1.2cm,align=center,draw,anchor=center},
    draw,inner sep=0]
    { \diagbox[width=\dimexpr \textwidth\relax, height=1.2cm]{Implementation}{PESQ version}     & {Narrowband raw PESQ (P.862) \\ 8 kHz / 16 kHz } & {Narrowband MOS-LQO (P.862.1) \\ 8 kHz / 16 kHz} & {Wideband MOS-LQO (P.862.2) 16 kHz \textbf{without} Corrigendum~2} & |[fill=yellow!40]| {Wideband MOS-LQO (P.862.2) 16 kHz \textbf{with} Corrigendum~2} & {Accepts \\ stereo inputs} \\ 
        {ITU reference code\\(ANSI C)} & \yesRMSE{0.62} & \yesRMSE{0.61} & |[fill=green!40]| {\ding{52} \\ (Comparison point)} & |[fill=red!40]| {\ding{56} \\ (RMSE = 0.56)} & \yesQM \\
        {ludlows/PESQ\\(Python)}   & \no & \yes & \yesRMSE{0.03} & \no & \no \\
        {audiolabs/torch-pesq\\(PyTorch)}   & \no & \no & \yesRMSE{0.15} & \no & \no \\
        {vBaiCai/python-pesq\\(Python)}   & \yes & \no & \no & \no & \no \\
        {Loizou\\(MATLAB)}   & \yes & \no & \no & \no & \no \\
        |[fill=yellow!40]| {audiolabs/PESQ\\(Python)}   & \no & \yes & \no & \yes & \no \\
    };
\end{tikzpicture}
\caption{Overview of the main PESQ versions and their publicly available implementations. The PESQ versions are listed in columns, with software implementations marked by a check (\ding{52}). Our audiolabs/PESQ provides wideband MOS-LQO (P.862.2) \textbf{with} Corrigendum~2, which is not implemented in the other openly available distributions.
Differences between versions can be substantial, as reported by the \ac{RMSE} with respect to the ITU reference implementation P.862.2 \textbf{without} Corrigendum~2.\label{tab1}}
\end{table*}

\subsection{Narrowband MOS-LQO (P.862.1)}
\begin{figure}[tb]
\centering
\centerline{\includegraphics[width=1\linewidth]{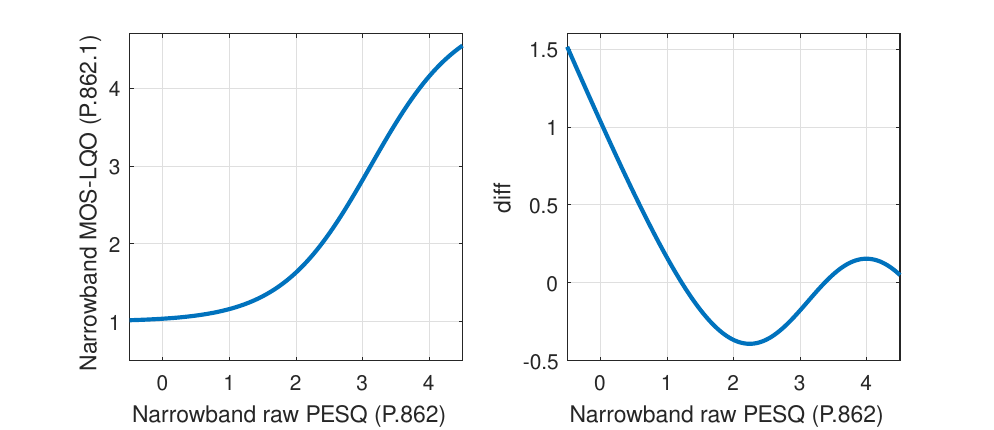}}
\caption{{\label{fig:map}}Mapping from narrowband raw PESQ (P.862) to narrowband \mbox{MOS-LQO} (P.862.1) in the left-hand figure and difference (diff) plot on the right.}
\end{figure}

A second version was recommended in P.862.1 \cite{PESQ:682.1}, in which raw PESQ scores are mapped to MOS-LQO (Mean Opinion Score - Listening Quality Objective \cite{ITUTP:800.1}), as shown in Fig.~\ref{fig:map}.
Both P.862 and P.862.1 support sampling frequencies of 8 and 16~kHz and assume narrow-band telephone-like listening conditions, i.e., with a strong attenuation below 300~Hz and above 3100~Hz.

\subsection{Wideband extension (P.862.2)}
The third version of PESQ is P.862.2 \cite{PESQ:682.2}. This considers a larger frequency range (50-7000~Hz, named \textit{wideband} or \textit{headphone listening}) and is available only for signals with a sampling frequency of 16~kHz.
This version only outputs MOS-LQO scores and no raw PESQ scores. 
Compared to P.862.1, P.862.2 introduces new input filters and a new output mapping to MOS-LQO.

\subsection{P.862 Corrigendum~2}

More than a decade after the first edition of P.862.2, the filter coefficients for wideband assessment were acknowledged to be incorrect in P.862 Corrigendum~2 \cite{PESQ:c2}.
The corrigendum describes a systematic under-prediction of subjective scores. The under-prediction of P.862.2 is estimated to be $0.8$ on average, and the corrected filter coefficients are provided.
Despite the acknowledged errors, P.862.2 without Corrigendum~2 seems to have established itself as the de facto standard in the speech processing community. This could be explained by the lack of an openly available implementation that directly includes Corrigendum~2. We fill this gap by introducing audioalabs/PESQ, as explained in Sec.~\ref{sec:py}.

\section{PESQ Implementations}
\label{sec:imp}

This section reviews the main openly available implementations of PESQ.

\subsection{ITU reference implementation (ANSI-C)}
The ITU provides a \textbf{reference ANSI-C} implementation\footnote{https://www.itu.int/rec/T-REC-P.862-200511-W!Amd2/en} that can output all main PESQ versions, with the exception of Corrigendum~2. In fact, Corrigendum~2 specifies these changes in text form, and the actual corrections are left to be implemented by the users of the reference code.

\subsection{Python and PyTorch}
\label{sec:py}
\hspace{1.5em}\textbf{ludlows/PESQ}\footnote{https://github.com/ludlows/PESQ} wraps around the ITU reference code and provides access to P.862.1 and P.862.2 for Python users. It is released as the PyPI package \textbf{pesq}.

\textbf{audiolabs/PESQ}\footnote{https://github.com/audiolabs/PESQ} is the repository we created by forking from ludlows/PESQ and adding the corrections detailed in Corrigendum~2. It is released as the PyPI package \textbf{pesqc2}. This provides the possibility of easily and reliably computing P.862.2 with Corrigendum~2.

\textbf{audiolabs/torch-pesq}\footnote{https://github.com/audiolabs/torch-pesq} provides a torch implementation of P.862.2 without Corrigendum~2 that can be readily used as loss function for training \acp{DNN}.

\textbf{vBaiCai/python-pesq}\footnote{https://github.com/vBaiCai/python-pesq} is also known as \textbf{pypesq} and only implements the narrowband raw PESQ (P.862). Therefore, this package is not suitable for users interested in the more popular wideband PESQ version (P.862.2).

\subsection{MATLAB}
A MATLAB implementation is available from the CD accompanying \textbf{Loizou}'s textbook \cite{Loizou:2007} (here we consider the 1st edition), and also online as obfuscated, execute-only code\footnote{https://ecs.utdallas.edu/loizou/speech/software.htm}. Both provide narrowband raw PESQ (P.862) only. Therefore, these packages are also not suitable if wideband PESQ (P.862.2) is of interest.

\section{Comparative Analysis}
\label{sec:differences}
This section quantifies the differences between the different versions (Sec.~\ref{sec:comp_v}) and the different implementations (Sec.~\ref{sec:comp_i}) of PESQ. The audio signals used in these comparisons are detailed in Sec.~\ref{sec:odaq}.

\subsection{Test audio signals}
\label{sec:odaq}
As test audio signals, we use the \ac{ODAQ} \cite{odaq}. It is an openly available dataset of stereo audio signals with $44.1$ or $48$~kHz sampling frequency. The dataset was designed to span a heterogeneous range of processing types and quality levels, i.e., from \textit{bad} to \textit{excellent} perceived quality. This is particularly useful for our analysis, as it allows us to analyze PESQ scores that span the entire quality scale.
For the evaluation in this section, the stereo signals are passively downmixed (i.e., by averaging the two channels) before computing PESQ. The signals are also downsampled to $16$~kHz using librosa~\cite{librosa} with its default parameters, i.e., soxr\_hq.

We consider the listening test part of \ac{ODAQ} that contains 240 audio samples with an average duration of 12.5 seconds. 
Of these, 80 samples contain speech panned to the phantom center mixed with stereo music, effects, and environmental sounds. Speech signals are processed with 5 different \ac{SE} systems, 4 of which are \ac{DNN}-based \ac{SE} systems from the literature. The \ac{SE} systems provide clearly distinct quality levels. The non-speech signals mainly include music and are processed by methods simulating quality degradations possibly encountered during audio coding.

Since the goal of this section is to examine how PESQ scores differ for the same input signals across versions and implementations, both speech and non-speech signals are considered. This ensures that PESQ mechanics are evaluated across a broad range of conditions and operating points, including conditions outside of the PESQ application domain. When computing PESQ, we use the reference of the listening test in \ac{ODAQ} as reference signal for PESQ.

ODAQ also provides ground-truth subjective scores of perceived quality. These are not used in this section. In contrast, they are used in Sec.~\ref{sec:stereo_quality}, but in that case only speech signals are considered.

\subsection{Comparing versions}
\label{sec:comp_v}
In this section, we evaluate the differences between PESQ versions based on their implementation in the ITU reference code. 
We report Pearson's linear correlation $\rho$, \acf{RMSE}, mean difference, and max absolute difference.
\acp{RMSE} are summarized in the first row of Table~\ref{tab1}.

\begin{figure}
\centering
\centerline{\includegraphics[width=1\linewidth]{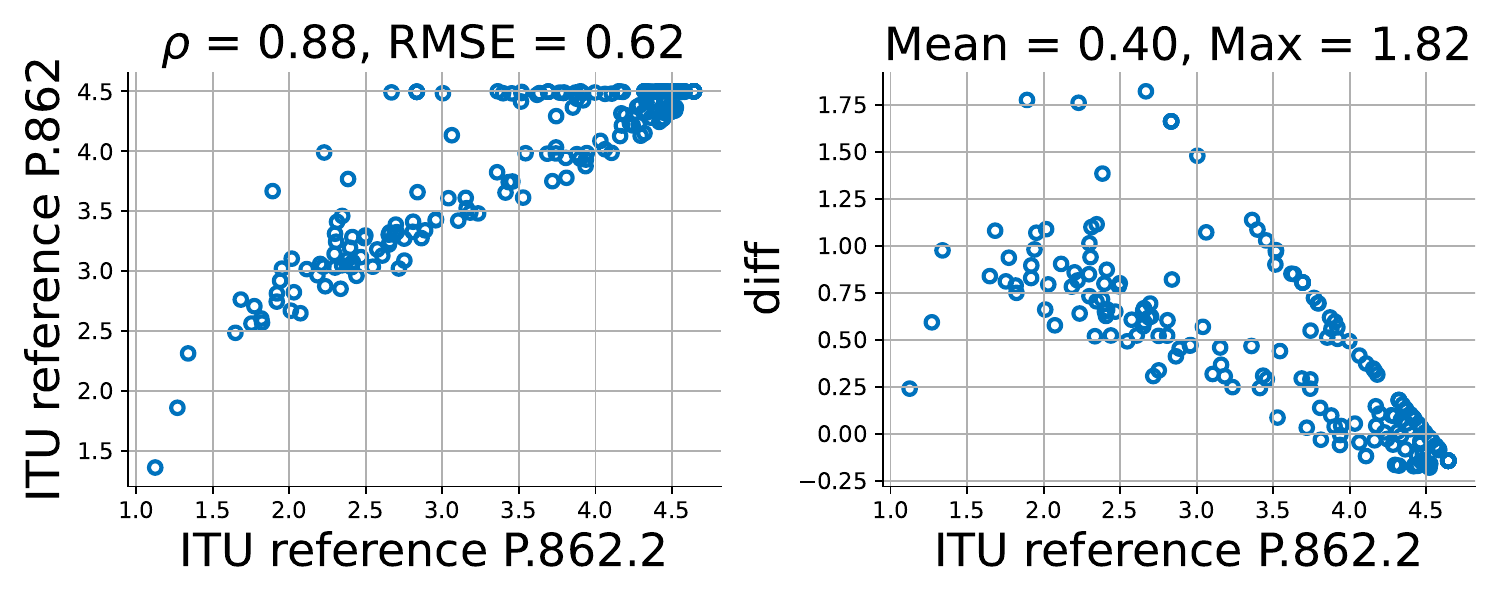}}
\caption{{\label{fig:raw}}Raw PESQ scores (P.862) compared with wideband PESQ (P.862.2 without Corrigendum~2). Reporting Pearson's correlation $\rho$, \ac{RMSE}, mean and max difference. 
}
\end{figure}

Figure~\ref{fig:raw} compares wideband P.862.2 without Corrigendum~2 to narrowband raw PESQ (P.862), using a sampling frequency of 16~kHz. The max difference is $1.82$ and \ac{RMSE}~$=0.62$ on the considered data.
The trend is similar (hence, the plot is not included) when comparing P.862.2 and P.862.1 (16~kHz), resulting in $\rho=0.89$, RMSE~$=0.61$, mean difference~$=0.41$, and max difference~$=0.92$.

\begin{figure}
\centering
\centerline{\includegraphics[width=1\linewidth]{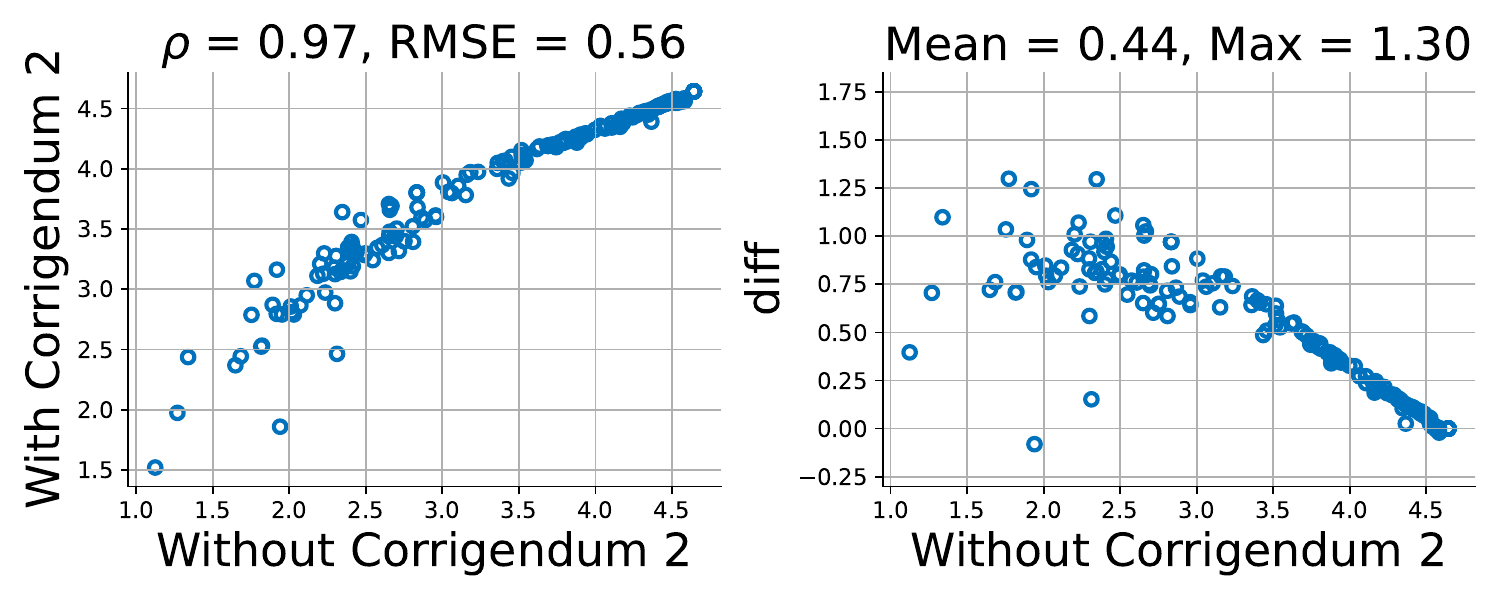}}
\caption{{\label{fig:c2}}PESQ P.862.2 with and without Corrigendum~2.
}
\end{figure}

Figure~\ref{fig:c2} shows the differences between P.862.2 with and without Corrigendum~2. Although the correlation is very strong, the differences in absolute terms are significant with an \ac{RMSE} $=0.56$ and a maximum difference of $1.30$ for the audio signals considered. 

\subsection{Comparing implementations}
\label{sec:comp_i}
\begin{figure}
\centering
\centerline{\includegraphics[width=1\linewidth]{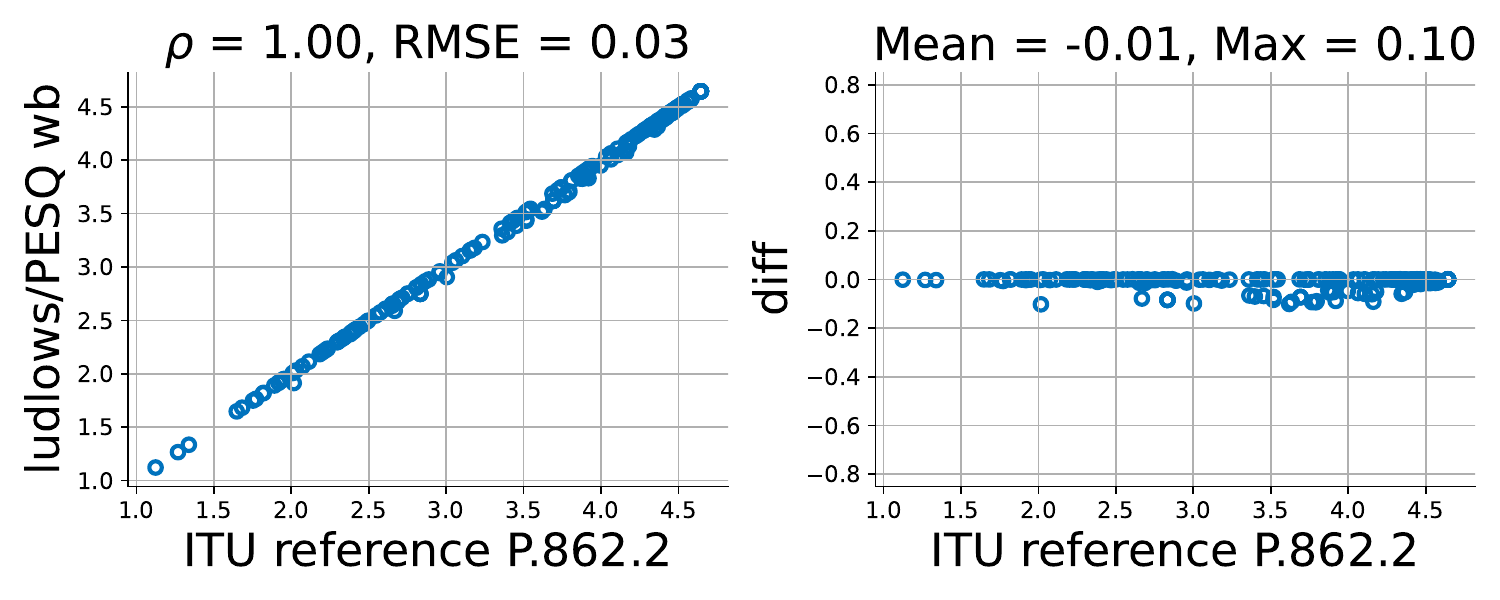}}
\caption{{\label{fig:ludlows}}ludlows/PESQ compared with the ITU reference implementation. Differences are negligible.}
\end{figure}

\begin{figure}
\centering
\centerline{\includegraphics[width=1\linewidth]{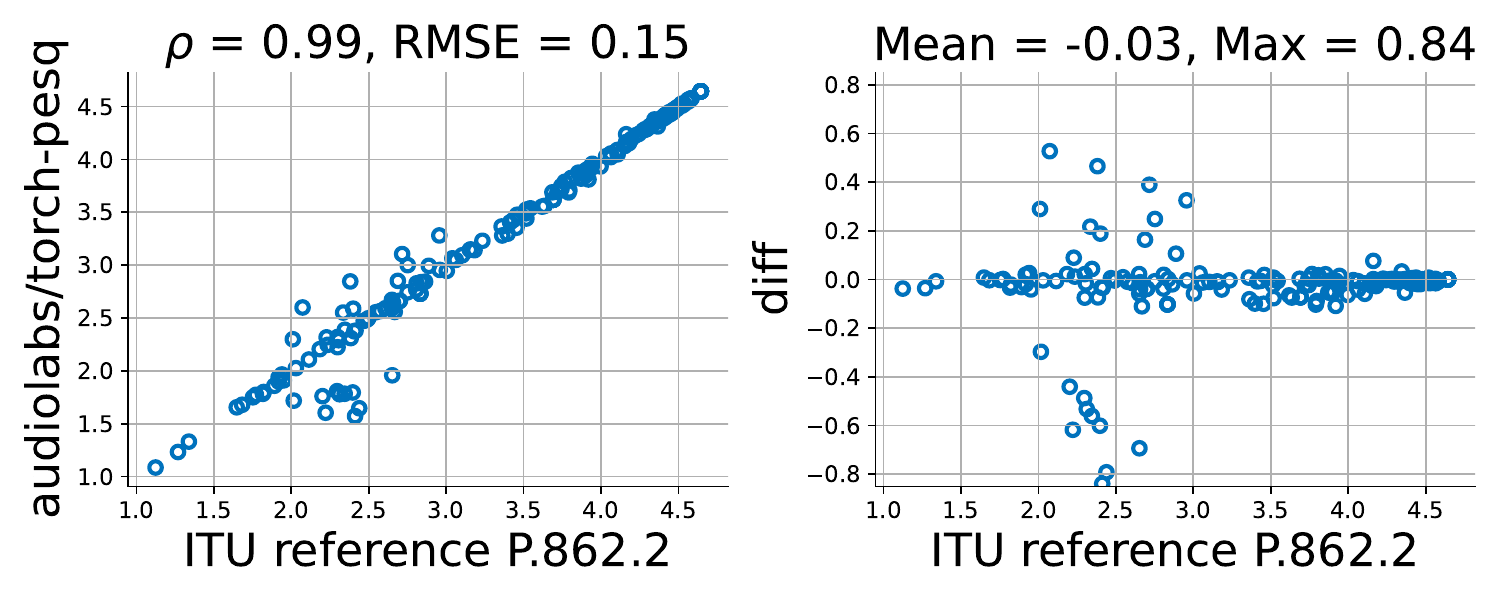}}
\caption{{\label{fig:torch}}audiolabs/torch-pesq compared with the ITU reference implementation.
}
\end{figure}

In this paragraph, we focus on P.862.2 (without Corrigendum 2) and compare different implementations. \acp{RMSE} are summarized in the third column in Table~\ref{tab1}.
Figure~\ref{fig:ludlows} compares P.862.2 as output by the ITU reference code and by ludlows/PESQ, showing that they are almost identical.
Figure~\ref{fig:torch} compares audiolabs/torch-pesq and the ITU reference implementation. As acknowledged by the authors of torch-pesq, some items can deviate significantly from the ITU reference implementation. This is because some processing steps were approximated to allow for a differentiable implementation of PESQ. Still, the average error is rather small with \ac{RMSE} $=0.15$ on the data considered.

Furthermore, our \mbox{audiolabs/PESQ} was compared against the ITU reference implementation with the addition of Corrigendum 2. The differences are negligible (hence, plot not included), with $\rho=1.00$, \ac{RMSE} $=0.01$, mean difference $=0.00$, and max absolute difference $=0.12$. This confirms that \mbox{audiolabs/PESQ} is a valid out-of-the-box solution for computing P.862.2 with Corrigendum 2.

\section{Multi-Channel}
\label{sec:multi}
\subsection{Reference implementation}
The results shown so far are obtained considering signals downmixed to mono.
Indeed, PESQ was developed for mono inputs, and all considered publicly available implementations accept only mono inputs (or batches thereof), with one important exception: The ITU reference code accepts multi-channel inputs.
It is not clear whether this is intentional. In the reference code, the reading routines load all input file samples. In practice, when a multi-channel audio file is given as input, a mono signal is created, in which channels are interleaved. As a consequence, the reference implementation processes the interleaved multi-channel signals as a long single-channel signal.

\subsection{Strategies for applying PESQ to stereo}
\label{sec:comp_s}

\begin{figure}
\centering
\centerline{\includegraphics[width=1\linewidth]{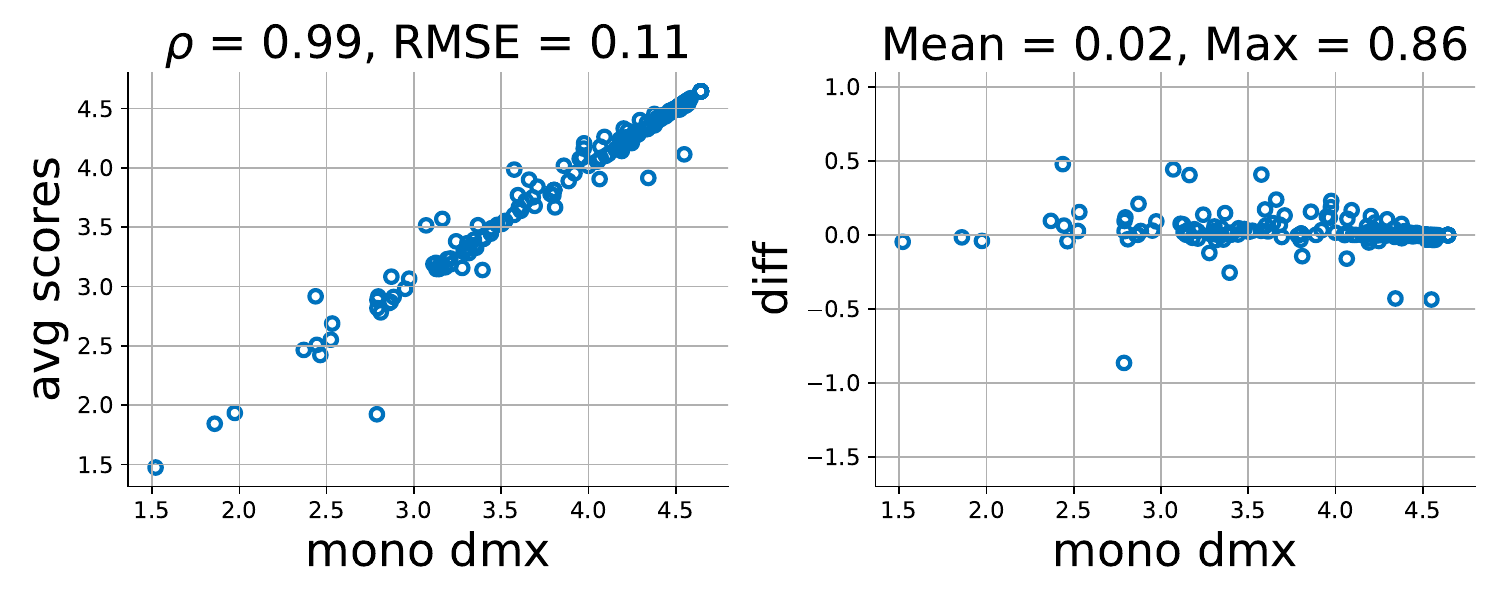}}
\caption{{\label{fig:avgch}}Comparing strategies for using PESQ (P.862.2 c2) with stereo signals: \textsc{mono dmx} and \textsc{avg scores}.}
\end{figure}

\begin{figure}
\centering
\centerline{\includegraphics[width=1\linewidth]{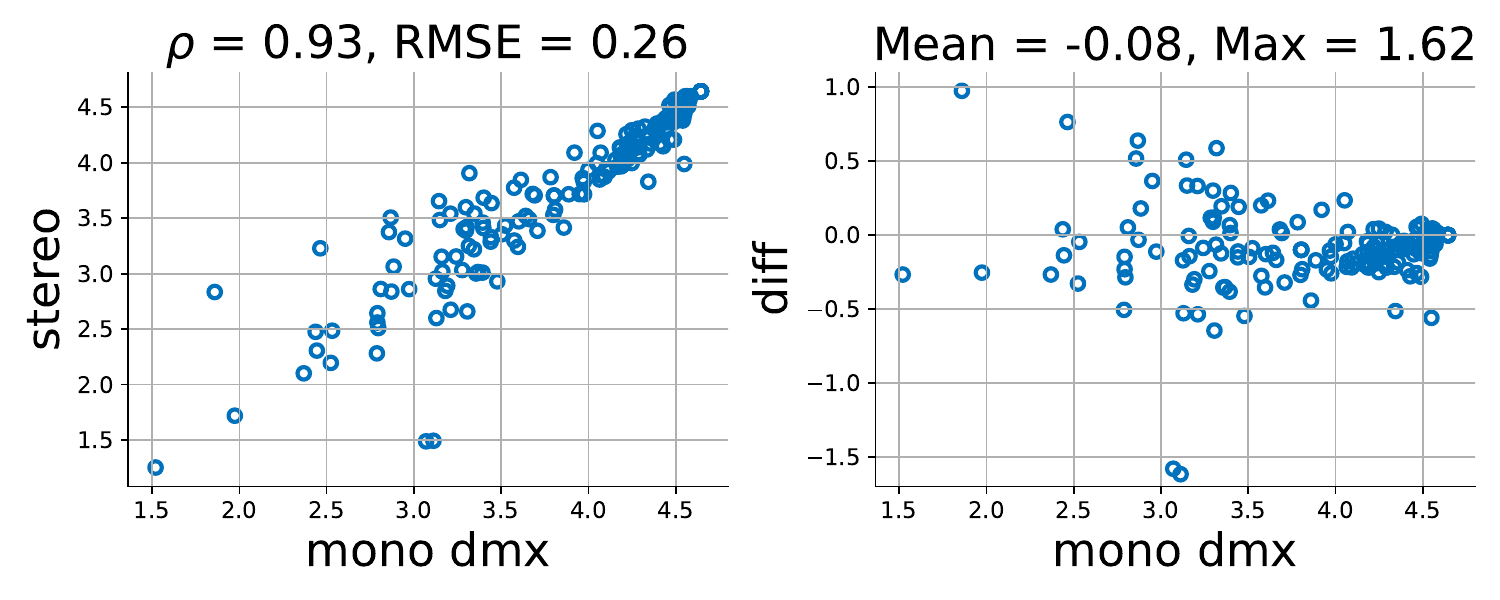}}
\caption{{\label{fig:mono}}Stereo signals input to the ITU reference implementation (with c2) against mono-downmixed inputs.}
\vspace{-1em}
\end{figure}

The processing of stereo signals by the ITU reference code is referred to as \textsc{stereo}.
More intuitive strategies for applying PESQ to stereo signals include:
\begin{enumerate}
    \item Mono-downmixing the stereo audio signal before computing PESQ (\textsc{mono dmx}).
    \item Evaluating PESQ for the individual channels and then averaging the resulting scores (\textsc{avg scores}). 
\end{enumerate}
These stereo strategies are compared in Fig.~\ref{fig:avgch} and Fig.~\ref{fig:mono}, in both cases using the ITU reference implementation with the addition of Corrigendum~2 (c2). 

\subsection{Comparing stereo strategies}
\label{sec:stereo_quality}
To better understand these strategies, we analyzed the correlation with subjective scores of perceived audio quality.
As the ground truth, we used the scores collected from two MUSHRA \cite{MUSHRA} listening tests involving stereo signals with speech panned to the phantom center of the stereo scene: 
\begin{enumerate}
    \item The \ac{DE} subset of \ac{ODAQ} containing only speech signals. A more detailed analysis for the correlation on this dataset is given in ~\cite{dick:2024}.
    \item The results from the listening test described in~\cite{strauss2021hands}, which compares \ac{SE} methods applied to real-world broadcast speech signals.
\end{enumerate}

We observed sensible PESQ scores in all cases, with Pearson's linear correlation coefficients $\ge 0.73$ and reaching $0.85$ in the best cases.
However, inconclusive results are observed, and no stereo strategy consistently outperforms the others on both considered datasets.
This prevents drawing a definitive conclusion on the overall best strategy.

\section{Conclusion}
\label{sec:conc}
Although replaced by POLQA, withdrawn by the ITU, and outperformed by other quality measures, PESQ remains a widely used speech quality metric. The term \textit{PESQ} encompasses different versions: narrowband raw PESQ score (P.862), narrowband MOS-LQO (P.862.1), and wideband MOS-LQO (P.862.2), with and without Corrigendum~2. 

This paper provides practical guidance on the different versions and implementations of PESQ. Differences between PESQ versions are shown to be particularly significant. Hence, it is recommended to specify the PESQ version and implementation when reporting PESQ scores, as this ensures clarity and reproducibility across studies.

In particular, Corrigendum 2 introduced significant adjustments to P.862.2, but it was not included in any of the publicly available implementations of PESQ. Probably due to this lack, the community seemed to ignore the corrigendum in practice, and P.862.2 without Corrigendum~2 is used most frequently.
Our audiolabs/PESQ fills this gap and provides easy and reliable access to P.862.2 with Corrigendum~2, i.e., the last version of PESQ.

Finally, attention is drawn to the likely unintentional behavior of the ITU reference implementation when dealing with multi-channel signals. Alternative strategies are discussed, providing plausible quality scores for stereo signals with speech panned to the center.

\balance
\small
\bibliographystyle{ieeetr}
\bibliography{literature}


\end{document}